\documentclass[11pt]{asaproc}

\usepackage{times}
\usepackage{fancyvrb}
\usepackage{relsize}
\usepackage{graphicx}
\usepackage{amsmath}
\usepackage{hyperref}

\begin{document}

\title{A New Framework for Random Effects Models}
 
\author{Norn Matloff \\ Dept. of Computer Science\\ University of
California, Davis}

% \date{November 9, 2015}
 
\maketitle

\begin{abstract}

A different general philosophy, to be called Full Randomness (FR), for
the analysis of random effects models is presented, involving a notion
of reducing or preferably eliminating fixed effects, at least formally.
For example, under FR applied to  a repeated measures model, even the
number of repetitions would be modeled as random.  It is argued that in
many applications such quantities really are random, and that
recognizing this enables the construction of much richer, more probing
analyses.  Methodology for this approach will be developed here, and
suggestions will be made for the broader use of the approach.  It is
argued that even in settings in which some factors are fixed by the
experimental design, FR  still ``gives the right answers.'' In addition,
computational advantages to such methods will be shown.

\end{abstract}

\section{Overview}

As a simple starting example, consider the classic random effects model
\cite{jiangbook}, with data $Y_{ij}$ modeled as

\begin{equation}
\label{classic}
Y_{ij} = \mu + \alpha_i + \epsilon_{ij}, ~ i = 1,...,r, j = 1,...,n_{i}
\end{equation}

for an unknown constant $\mu$ and with $\alpha_i$ and $\epsilon_{ij}$
modeled at random variables having mean 0 and variances $\sigma_a^2$ and
$\sigma_e^2$ respectively.  These random variables are assumed
independent across $i$ and $j$, though assumptions will generally not be
made here about their distributions.

The present paper advocates treating quantities such as the $n_{i}$ as
random variables, using a capital letter to emphasize this, $N_{i}$:

\begin{equation}
\label{nrand}
Y_{ij} = \mu + \alpha_i + \epsilon_{ij}, ~ i = 1,...,r, j = 1,...,N_{i}
\end{equation}

As seen below, we will also treat regressor variables, if any, to be
random.

In short, the goal of this paper is to  encourage analysts to model all
quantities as random, even in most cases those fixed by experimental
design.  Advantages to this approach will turn out to include:

\begin{itemize}

\item Much richer, more probing analyses can be devised.

\item The derivation of estimators and their standard errors can be
simplified.

\item For some large problems, the computation for fully random models
can be parallelized, under a method known as Software Alchemy.

\end{itemize}

% For the time being, we will make the same assumptions as before, e.g.\
% that $\alpha_i$ has mean 0 and variance $\sigma_a^2$, conditional on
% $N_{ij}$.

\section{Advantages of Treating the Ordinarily-Fixed Quantities As Random}

Let's begin with the $N_i$. Why model them as random?  

As a motivating example of the topic here, consider recommnder systems
\cite{ricci}, such those that might be applied to the Movie Lens data
\cite{herlocker}, with ratings of many movies by many users.  If one
views this ratings data in matrix form, as do Gao and Owen
\cite{gaoowen} and Perry \cite{perry},, with rows and columns corresponding
to users and movies, respectively, then the matrix is sparse:  In the
notation of \cite{gaoowen}, $z_{ij} = 0$ for most $i$ and $j$, where
$z_{ij}$ is an indicator variable for whether user $i$ has rated movie
$j$.  The authors in that paper consider the users to be a random sample
from the potential population of all users, and similarly for the
movies, and thus use a random effects model.

Details on that model will be presented shortly, but for now, let's
consider only the users, not the movies.  Then we might model the data
using (\ref{classic}) or (\ref{nrand}), with $\sigma_a^2$ being a
measure of ratings variability from user to user.

Our FR approach might be used, for instance, if we suspect that users
who rate a lot of movies become jaded, thus tending to give lower
ratings.  In other words, there may be a statistical relation between
$N_i$ and $\alpha_i$.  If such a relation were estatblished, we may wish
to discount the ratings of users having large $N_i$.\footnote{Actually,
there is negative relation like this for the Movie Lens data, with the
quantities $\sum_{j=1}^{N_i} Y_{ij} / N_i$ having a
statistically significant but small relation to $N_i$ } 

To investigate this, it is natural to model the $N_i$ as having their
own effects, just as we do for the $\alpha_i$, say with a model

\begin{equation}
\label{famsize}
Y_{ij} = c_1 + c_2 N_i + \alpha_i + \epsilon_{ij}, ~ i = 1,r, j = 1,...,N_{i}
\end{equation}

The quantities $\alpha_i \textrm{ and } \epsilon_{ij}$ are now assumed
independent conditional on $N_i$, and their variances, $\sigma_a^2$ and
$\sigma_e^2$, are now conditional on $N_i$ as well.  The $N_i$ are
considered i.i.d.  The quantity $\alpha_i$ now represents the overall
rating tendency for user $i$, after the effect of count of ratings has
been removed.

The modeling of the $N_i$ as a variance component could be useful in
many different application fields.  It is known, for example, that there
is a negative correlation between family size and household income
\cite{berger}.  If the observation units in a study are children within
families, it would be thus useful to incorporate the number of children
$N_i$ into the analysis.  A study of workers at various companies may be
similar to this.

% \subsection{Treating Regressors As Random}

It is common to include linear-model terms into (\ref{classic}):

\begin{equation}
\label{randomx}
Y_{ij} = x'\gamma  + \alpha_i + \epsilon_{ij}, ~ i = 1,r, j = 1,...,n_{i}
\end{equation}

for a vector of known regressors $x$ and unknown constant vector $\gamma$.
(Our old term $\mu$ is now folded in by inserting a 1 element in $x$.)
But it may be helpful to consider the regressors random also, so that
our model becomes

\begin{equation}
\label{xigamma}
Y_{ij} = X_i'\gamma  + \alpha_i + \epsilon_{ij}, ~ i = 1,r, j = 1,...,N_{i}
\end{equation}

where again the use of a capital letter indicates a random variable,
with the $X_i$ i.i.d.

In (\ref{randomx}), we may even wish to reverse the usual prediction
relationship, predicting one or more of the regressors from the
$Y_{ij}$.  In the case of recommender systems, for example, the analyst
may wish to infer certain information about the user.  Some values of
the regressors may be missing, for instance, and we may wish to impute
them using the other variables.  This would be even more reason to treat
the regressors as random.

% Similarly, consider the {\bf InstVal} data included with the R
% random-effects modeling package {\bf lme4}.  It's essentially another
% recommender system, with university students rating their instructors.
% Presumably the higher-quality instructors are chosen by more students,
% and a valuable approach to studying this could be to try to predict the
% $N_i$ by the $Y_{ij}$.
% 
% In this manner, very elaborate, powerful models can be developed.  For
% instance, the FR approach can be used to extend and enhance popular
% recommender systems methodology such as {\it collaborative filtering}. 

\section{Multicomponent Models}

The approach can also be used in models with more than one
variance component.  For instance, consider the model used by
\cite{gaoowen} with the movie data,

\begin{equation}
\label{movie}
Y_{ij} = \mu + \alpha_i + \beta_j + \epsilon_{ij}, ~
i = 1,...,r, ~ j = 1,...,c, ~ z_{ij} = 1
\end{equation}

Here $r$ is the number of users and $c$ is the number of movies.

Applying our method to this model, we treat the $z_{ij}$ as random
variables $Z_{ij}$, 

\begin{equation}
\label{nrand1}
Y_{ij} = \mu + \alpha_i + \beta_j + \epsilon_{ij}, ~
i = 1,...,r, ~ j = 1,...,c, ~ Z_{ij} = 1
\end{equation}

and define the row and column observation counts,

\begin{equation}
N_i = \sum_{j=1}^c Z_{ij}
\end{equation}

\begin{equation}
M_j = \sum_{i=1}^r Z_{ij}
\end{equation}

The $N_i$ and $M_j$ are then random as before.

We might also bring in random regressors, for both users and movies:

\begin{equation}
Y_{ij} = \mu + U_i'\gamma +V_j'\eta +
\alpha_i + \beta_j + \epsilon_{ij}, ~
i = 1,...,r, ~ j = 1,...,m, ~ Z_{ij} = 1
\end{equation}

\section{Estimation Methodology}

The Method of Moments (MM) is an attractive approach here, as it will
enable estimation of, for instance, $\sigma_a^2$ in (\ref{nrand})
without assuming a particular distribution family for the $\alpha_i$
\cite{cheng} \cite{milliken} \cite{gaoowen}. 

Let's take (\ref{nrand}) as our example, using 

\begin{equation}
\label{yidot}
Y_{i.} = \sum_{j=1}^{N_i} Y_{ij}, ~~ i = 1,...,r
\end{equation}

as our pivot quantity.  It will be very helpful to define generic
versions of the variables:  Let $Y$, $N$, $\alpha$, $\epsilon$ and $S$
have the same distributions as $Y_{ij}$, $N_i$, $\alpha_i$,
$\epsilon_{ij}$ and $Y_{i.}$.

Also, let $\epsilon_j, ~ j = 1,2,...$ be i.i.d.\ with the distribution of 
$\epsilon$.  Then write 

\begin{equation}
\label{seqn}
S = N \mu + N \alpha 
+ \epsilon_1 + ... 
+ \epsilon_N
\end{equation}

Now apply the ``Pythagorean Theorem for Expectations,'' 
 
\begin{equation}
Var(U) =
E \left [ Var(U |V) \right ] +
Var \left [ E(U |V) \right ]
\end{equation}

to (\ref{seqn}).  First,

\begin{eqnarray}
E \left [ Var(S ~|~ N) \right ]
&=& E \left [ N^2 \sigma_a^2 + N \sigma_e^2 \right ] \\ 
&=& (\nu_2+\nu_1^2) \sigma_a^2 + \nu_1 \sigma_e^2
\end{eqnarray}

where $\nu_1 \textrm{ and } \nu_2$ are the population mean and variance
of $N$.  

Next,

\begin{equation}
Var \left [ E(S ~|~ N) \right ]
= \mu^2 \nu_2
\end{equation}

In other words,

\begin{equation}
\label{vars}
Var(S) = (\nu_2+\nu_1^2) \sigma_a^2 + \nu_1 \sigma_e^2 + \mu^2 \nu_2 
\end{equation}

% Similarly, the distribution of 
% 
% \begin{equation}
% Y_{..} = \sum_{i=1}^r \sum_{j=1}^{N_i} Y_{ij}
% \end{equation} 
% 
% is the same as that of
% 
% \begin{equation}
% M \mu +
% M \alpha + 
% \epsilon_1+...+\epsilon_M
% \end{equation}
% 
% where
% 
% \begin{equation}
% M = N_1+...+N_r
% \end{equation}
% 
% So, reasoning as above
% 
% \begin{equation}
% \label{varydotdot}
% Var(Y_{..}) = 
% (r \nu_2 + r^2 \nu_1^2) \sigma_a^2 + 
% r \nu_1 \sigma_e^2 + r \mu^2 \nu_2
% \end{equation}

Also, 

\begin{equation}
\label{vary}
Var(Y) = \sigma_a^2 + \sigma_e^2
\end{equation}

We have 5 unknowns to estimate --- $\sigma_a^2$, $\sigma_3^2$, $\mu$,
$\nu_1$ and $\nu_2$ --- and thus need 5 equations for MM.  (\ref{vars})
and (\ref{vary}) provide the right-hand sides of 2 equations, with the
left-hand sides being the sample variances of the $Y_{i.}$ and the 
$Y_{ij}$, respectively.  The other 3 equations come quite simply:  We
estimate the $\nu_m$ by the sample mean and variance of $N$, and
estimate $\mu$ by $Y_{..} / M$, where $M = N_1+...+N_r$.

The estimation of more advanced models can be approached similarly,
i.e.\ deriving expressions for variances and means, typically with the
aid of the ``Pythagorean Theorem.''  

In the regression setting (\ref{xigamma}), since we have

\begin{equation}
EY_{ij} = X_i' \gamma
\end{equation}

we can estimate $\gamma$ separately using standard linear model methods,
and proceed as before.  

Note, though, that with the FR approach, MM equations may be
nonlinear.  For example, consider (\ref{famsize}).  The details will not
be presented here, but the key points are as follows:  The term $N \mu$
in (\ref{seqn}) now becomes

\begin{equation}
\label{nneqn}
N (c_1 + c_2 N) = c_1 N + c_2 N^2
\end{equation}

Taking the variance of this quantity then brings in the third and fourth
moments of $N$, and produces product terms such as $c_1 c_2$.  The
former issue is no problem, as the moments are readily estimated from
the $N_i$, but the latter issue means we are now dealing with nonlinear
equations in the parameters to be estimated. Computation then must be
done iteratively.

It is convenient to not write explicit expressions for the variance of
(\ref{nneqn}), but simply write

\begin{equation}
\label{newvarn}
Var(c_1 N + c_2 N^2)
\end{equation}

At each iteration, we take our current estimates of the $c_k$, and
compute the sample variance of the quantities

\begin{equation}
c_1 N_i + c_2 N_i^2
\end{equation}

as our estimate of (\ref{newvarn}).

\section{What If the Quantities Are Not Random?}

In many applications of random effects models, quantities such as $n_i$
and $x_i$ above are fixed in the experimental design.  However, one can
show that typically the same estimators emerge, whether one assumes a
random $N_i$ or fixed $n_i$.  The same is true for regressors.

As a quick example, consider (\ref{classic}).  Instead of
(\ref{vars}), we have

\begin{equation}
Var(Y_{i.}) = n_i^2 \sigma_a^2 + n_i \sigma_e^2
\end{equation}

Also, $EY_{i.} = n_i \mu$.

Say we set up MM by equating the sample average of the $Y_{i.}^2$ to its
expectation.  The latter would be

\begin{equation}
\frac{1}{r} 
\sum_{i=1}^r [ Var(Y_{i.}) + (EY_{i.})^2 ] =
\frac{1}{r} 
\sum_{i=1}^r 
[n_i^2 \sigma_a^2 + n_i \sigma_e^2
+ (n_i \mu)^2]
\end{equation}

Even without algebraic simplification, it's clear that the result will
be essentially the same as that obtained for the random $N_i$ model.
For instance, the term

\begin{equation}
\frac{1}{r} 
\sum_{i=1}^r 
n_i^2 \sigma_a^2
\end{equation}

corresponds to the term

\begin{equation}
(\nu_2 + \nu_1^2) \sigma_a^2
\end{equation}

in (\ref{vars}).  In essence, the above derivation is implicitly
treating the (constant) row counts as random, having a uniform distribution 
on $\{n_1,...,n_r\}$.

The significance of this is that one can enjoy the benefits of the FR
approach (Sections \ref{simplified} and \ref{sa}) even if the quantities
truly are fixed.

\section{Advantager opf FR: Simplified Derivation of MM Equations}
\label{simplified}

Equations in random effects analysis can become quite complex.  Note
for instance the conditions needed merely to establish consistency in
\cite{jiangjasa}.

This complexity certainly includes the settings of MM estimation.  For
instance, even in the simplest model, (\ref{vars}) seems rather
complicated in its form here, but is even more sprawling if the $n_i$
are taken as fixed.  We argue here that our FR method can greatly ease
the derivation of the MM equations.

This is especially true in light of our use of generic variables, as in
(\ref{seqn}), which can reduce large amounts of equation clutter.
Consider for example the model (\ref{movie}).  Suppose we need to find
the covariance between $Y_{k.}$ and $Y_{m.}$.  Once again, the details
will not be shown here, but a glance at (\ref{seqn}) shows that when we
will apply the covariance form of the ``Pythagorean Theorem,'' the key
quantity will be distributed as

\begin{equation}
\beta_1 + ... + \beta_T
\end{equation}

where $T$ is the number of columns that rows $k$ and $m$ have in common.
The distribution of $T$ can be estimated empirically, as we did for $N$
above.  The point is that all this can be done without any explcit
writing of the $Z_{ij}$.  The difference in complexity of expressions
between the FR and fixed-$z_{ij}$ approaches will be quite substantial.

\section{Computational Benefits}
\label{sa}

In random-effects modeling applications involving very large data sets,
a major concern is computation time and space.  As noted in
\cite{gaoowen}, the REML method of estimation in a two-component model,
for example, requires $O(d^3)$ time and memory space, where $d$ would be
either $r$ or $c$ in the movie ratings example above.  Indeed,
\cite{pennell} reported that ``SAS PROC MIXED ran out of memory when we
attempted to fit a model with random smoking effects.''

A method that I call Software Alchemy \cite{matloffsa} can help remedy
both time and memory problems in contexts of i.i.d.\ data,, using a very
simple idea.  Say we are estimating a population value $\theta$,
typically vector-valued.  One breaks the data into $g$ approximately
equal-size chunks, finds $\widehat{\theta}$ on each one, and then takes
the one's overall estimate to be

\begin{equation}
\overline{\theta} = \frac{1}{g} 
\sum_{i=1}^g \widehat{\theta}_i
\end{equation}

This changes the original problem into an ``embarrassingly parallel''
computational problem, i.e.\ easy to compute in parallel, say on $g$
machines in a cluster or on $g$ cores in a multicore machine.

This speeds up computation by a factor of $g$, and since each
$\widehat{\theta}$ requires only $1/g$ of the memory space requirement,
the method may remedy memory limitation problems in cluster settings.
In fact, the same is true even on a single-core machine, since one would
still need only $1/g$ of the memory space requirement at each iteratioo.

The procedure also gives us a mechanism for empirical computation of
standard errors.

It is shown in \cite{matloffsa} that if $\widehat{\theta}$ is
asymptotically normal, then the same will be true for
$\overline{\theta}$, and moreover, the latter will have the same
asymptotic covariance matrix as the former.  Thus no statistical
efficiency is lost.

The point then is that this can be applied profitably to random-effects
models --- if the i.i.d.\ requirement of Software Alchemy is
satisfied.\footnote{In \cite{matloffsa}, it is remarked that the theory
could be extended to the context of independent but nonidentically
distributed observations.  However, this would necessitate defining
complex application-specific conditions, plus the determination of
proper weights. All of this may be infeasible in complex random-effects
applications.} By making quantities like the $N_i$ random, this can be
done in many cases.

Consider the model (\ref{nrand}), for instance.  A set of key quantities
in the estimation procedure consists of the $Y_{i.}$.  By modeling the
$N_i$ as i.i.d., the same will be true for the $Y_{i.}$, and Software
Alchemy can used.

Now consider (\ref{nrand1}), a more subtle setting. Let $W_1$, $W_2$...
denote the $Y_{ij}$, arranged in the order in which the ratings are
submitted, and write

\begin{equation}
W_m = \mu + \alpha_{I_m} + \beta_{J_m} + \epsilon_m, m = 1,2,...
\end{equation}

where the $I_m$ and $J_m$ are now drawn in an i.i.d.\ manner from
distributions on {1,...,r} and {1,...,c}.  Assuming that submissions
come in to the rating site in an i.i.d.\ manner, this structure is
reasonable.  We can then divide the $W_m$ into chunks, estimate $\mu$,
$\sigma_a^2$ and $\sigma_e^2$ as before on each chunk, then average over
chunks.

\section{Relation to Mixing Distributions}

Note that random effects models can be viewed in terms of mixing
distributions, with the advantage, for example, that the entire
distribution of $\alpha$ might be estimated, rather than just its
variance \cite{mack} \cite{hakobyan}.  This might be used to develop
prediction intervals, say for a continuous $Y$.

\section{Conclusions}

This paper has presented Full Randomness, a proposed framework for the
enhanced analysis of random effects.  FR enables the formation of richer
models of the phenomena under study, simplifies derivations of complex
models, and can facilitate parallel speedup of computation.
Many further directions in methodology could be explored under this
framework, with applications to a number of specific fields, such as the
aforementioned collaborative filtering.  

{}

\end{document}